\documentclass[sigconf]{acmart}

\usepackage{lipsum}
\usepackage{multirow}
\usepackage{todonotes}
\usepackage{color, colortbl}
\definecolor{gray}{gray}{0.9}
\usepackage{tcolorbox}
\usepackage{enumitem}
\usepackage{subcaption}
\hypersetup{draft}

\AtBeginDocument{%
  \providecommand\BibTeX{{%
    \normalfont B\kern-0.5em{\scshape i\kern-0.25em b}\kern-0.8em\TeX}}}

\copyrightyear{2020} 
\acmYear{2020} 
\setcopyright{none}
\acmConference[ICGSE '20]{15th IEEE/ACM International Conference on Global Software Engineering}{October 5--6, 2020}{Seoul, Republic of Korea}
\acmBooktitle{15th IEEE/ACM International Conference on Global Software Engineering (ICGSE '20), October 5--6, 2020, Seoul, Republic of Korea}
\acmPrice{15.00}
\acmDOI{10.1145/3372787.3390436}
\acmISBN{978-1-4503-7093-6/20/05}

\begin{document}

\title[A Case Study on Tool Support for Collaboration in Agile Development]{A Case Study on Tool Support for Collaboration in\\Agile Development}

\author{Fabio Calefato}
\orcid{0000-0003-2654-1588}
\affiliation{%
  \institution{University of Bari}
  \streetaddress{Dipartimento di Informatica, via E. Orabona, 4}
  \city{Bari}
  \state{Italy}
  \postcode{70125}}
\email{fabio.calefato@uniba.it}

\author{Andrea Giove}
\affiliation{%
  \institution{University of Bari}
  \streetaddress{Dipartimento di Informatica, via E. Orabona, 4}
  \city{Bari}
  \state{Italy}
  \postcode{70125}}
\email{a.giove16@studenti.uniba.it}

\author{Filippo Lanubile}
\orcid{0000-0003-3373-7589}
\affiliation{%
  \institution{University of Bari}
  \streetaddress{Dipartimento di Informatica, via E. Orabona, 4}
  \city{Bari}
  \state{Italy}
  \postcode{70125}}
\email{filippo.lanubile@uniba.it}

\author{Marco Losavio}
\affiliation{%
  \institution{Klopotek Software \& Technology Services Italia}
  \city{Gioia del Colle}
  \country{Italy}
  }
\email{m.losavio@klopotek.it}

\renewcommand{\shortauthors}{Calefato et al.}

\begin{abstract}
We report on a longitudinal case study conducted at the Italian site of a large software company to further our understanding of how development and communication tools can be improved to better support agile practices and collaboration. 
After observing inconsistencies in the way communication tools (i.e., email, Skype, and Slack) were used, we first reinforced the use of Slack as the central hub for internal communication, while setting clear rules regarding tools usage. 
As a second main change, we refactored the Jira \textsc{Scrum} board into two separate boards, a detailed one for developers and a high-level one for managers, while also introducing automation rules and the integration with Slack. 
The first change revealed that the teams of developers used and appreciated Slack differently with the QA team being the most favorable and that the use of channels is hindered by automatic notifications from development tools (e.g., Jenkins). The findings from the second change show that 85\% of the interviewees reported perceived improvements in their workflow. \\
Despite the limitations due to the single nature of the reported case, we highlight the importance for companies to reflect on how to properly set up their agile work environment to improve communication and facilitate collaboration.
\end{abstract}



\maketitle

\newcommand{\klop}{\textsc{KSTS-It}}

\section{Introduction}
Agile development is an increasing trend for software organizations, whether small or large, working co-located or geographically distributed \cite{Papatheocharous2013}. 
Still, there are many challenges in supporting effective collaboration and communication for agile teams \cite{Lous2017-icgse,Lous2018-icssp,Lous2018-icgse}. 
For example, an ever-growing plethora of different tools are needed to develop and manage software projects \cite{Boehm2006,Calefato2016-hns}, placing teams into a situation of information fragmentation \cite{Storey2017} and overload \cite{Murphy2009}.

This paper describes a longitudinal, single case study conducted at the Italian site of a large, globally-distributed company named Klopotek.
Our research aims at studying the Agile work environment at Klopotek to identify points of friction between development practices and collaboration tools. As such, we are interested in answering the following research question:
\textit{How can the Agile work environment be improved to facilitate communication and better support the collaboration workflow?}

The case study lasted for about four months (from December 2018 to April 2019) and was executed into two consecutive steps. First, we gained an understanding (via direct observation and semi-structured interviews) of the work environment, that is, how development and collaborative tools are used. Then, in line with action research guidelines \cite{Avison1999,dosSantos2011}, we introduced  two main changes aimed respectively at improving the communication and collaboration workflow by optimizing how the Slack and Jira tools were used.
The first change revealed that the teams of developers used and appreciated Slack differently, with the QA team being the most favorable and that the use of channels is hindered by automatic notifications from development tools (e.g., Jenkins). The findings from the second change show that 85\% of the interviewees reported perceived improvements in their workflow.

The research presented here is first step focused on streamlining communication and collaboration at the Italian site of the company, before bringing and analyzing the changes to the other remote sites.
As main contributions, in our paper we provide \textit{(i)} recommendations on how  to set up a Slack workspace for supporting Agile teams as well as  an understanding of the (side) effects of \textit{(ii)} integrating notifications from external development tools with channels and \textit{(iii)} using multiple Jira boards with different perspectives and automation rules. 

The remainder of the paper is organized as follows. In Sect.~\ref{sec:case_description}, we describe the case subject and the settings for the direct observation and changes. In Sect.~\ref{sec:findings} and \ref{sec:discussion}, respectively, we report and discuss the results from our analyses, comparing our findings to related work. Finally, we conclude in Sect.~\ref{sec:conclusions}.

\section{Case Study}\label{sec:case_description}

In this section, we report about the mixed-method, longitudinal case study conducted on-site, following the guidelines provided by \citeauthor{Runeson2012}~\cite{Runeson2009,Runeson2012}.

Accordingly, below we first introduce the company (Sect.~\ref{sec:org}). Then, we describe the data collection procedure used to gather a clear understanding of the work environment and tooling (Sect.~\ref{sec:data}). Finally, we detail the two changes introduced \textit{in vivo} to improve communication and collaboration (Sect.~\ref{sec:changes}).

\subsection{Klopotek}
Founded in 1992 in Berlin, Germany, Klopotek AG is now a global company operating in the area of publishing software, with sites in Europe and the USA. Klopotek works with more than 350 publishers worldwide, for a total of over 14,000 users in 140+ locations.

In 2015, the company opened near Bari, Italy, a site dedicated to development. \textit{Klopotek Software \& Technology Services Italia} (\klop{}) was opened with the specific goal of redeveloping the legacy ERP desktop application into a cloud-based web application, called \textsc{Stream}. The other two projects on which \klop{} works are \textsc{Klopotek Deployment Manager} (KDM), a tool used both internally and by customers to install and configure Klopotek solutions without manual intervention, and \textsc{Core}, a framework that allows both the web app and the legacy tool to interact with the back-end via a RESTful API.

\subsubsection{Organization} \label{sec:org}
\klop{} is currently structured into five main areas, each corresponding to a specific team:
\begin{itemize}
    \item \textbf{\textit{Architecture}}. A team of senior architects in charge of the design and development of the new \textsc{Core} framework, while also taking care of the coexistence with the legacy solution.
    \item \textbf{\textit{Dev}}. The team consists of seven developers taking care of all the development and maintenance activity on some \textsc{Stream} products catalog. Some of them also contribute  to   \textsc{Core}.
    \item \textbf{\textit{QA}}. The team consists of three members, also associated with the Dev team, who take care of functional and non-functional testing, documentation,  as well as  process and product monitoring.
    \item \textbf{\textit{DevOps}}. A team of software engineers who work closely with Dev, QA, and Architecture, respectively, to speed up the integration/deployment of new features/releases and the configuration of the testing and development environments. 
    \item \textbf{\textit{Management}}. This board includes the \klop{} director
    , the leaders of each of the previous four teams as well as the Product Manager and the Technical Product Manager.
\end{itemize}

The \klop{} site is located in a three-floor building. 
Due to the co-location in one large room, within-team interaction typically happens face-to-face (F2F), and developers sometimes decide to pair-program, if needed, as well as testers who often work in pairs with developers. 

\subsubsection{Development Process}
The Dev and QA teams follow a hybrid process that is transitioning towards the \textsc{Scrum} methodology \cite{Schwaber2002}. Instead, at the time of the case study, the Architecture and DevOps team had not adopted any specific methodology, but were considering the same idea of transitioning to \textsc{Scrum} as well. 

Overall, the development process involves three main types of activities: \textit{analysis \& design}, \textit{development \& test}, and \textit{release \& delivery}. 


The \textit{development \& test} activities involve periodic Grooming and Sprint Planning sessions, organized typically on Monday afternoon or Tuesday morning. Here participants (i.e., the Dev team, the Dev team leader, and the Technical Product Manager) make estimations of user stories based on the development time (in person-days). 
Story points are assigned to user stories using the \textit{Planning Poker} technique \cite{Cohn2005,Calefato2011-ppoker}. 

Besides Grooming and Sprint Planning, also Daily and Sprint Review Meetings are conducted. The Daily Meetings are organized every day at 9:15, involving the Technical Product Manager and the team members of the Dev and QA teams (including the leaders of both teams). As for the Sprint Reviews, they also involve members from the Berlin site for better coordination and are followed by Retrospective Meetings, which instead involve only developers and testers, to give them the maximum freedom of expression. Here, the QA team is also responsible for assessing the quality of Sprints. 

The activities of \textit{release \& delivery} change depending on the product but typically involve \textit{Snapshot} releases updated daily, \textit{Beta} versions released at the end of each Sprint, \textit{Release Candidate}, i.e., the version under test before delivery, and \textit{Final}, i.e., the delivered version.


\begin{table}[t]
 \setlength{\abovecaptionskip}{-1pt}
  \caption{Tools used at \klop{}, arranged by category.}
  \label{tab:tooling}
  \footnotesize
  \begin{tabular}{lll}
    \toprule
\multicolumn{2}{l}{\textbf{Category}}                                                            & \textbf{Tool}                                                                                      \\ \midrule
\multirow{10}{*}{Development} &  SCM & GitLab                                                                                           \\
 & Coding                                                                  & \begin{tabular}[l]{@{}l@{}}Balsamiq, PapDesigner,
                                                                            \\Visual Paradigm, Eclipse,\\WebStorm, VS Code\end{tabular}
                                    \\
 & Issue Tracking                                                               & Jira
                                \\
 & Build                                                                        & Maven, Grunt                                                                                        \\
 & Testing                                                                      & \begin{tabular}[l]{@{}l@{}}JUnit4, JMeter,\\HP Quality Center\end{tabular}
                                \\
 & Inspection                                                                   & SonarQube 
                                    \\
 & Documentation                                                                & Confluence
                                     \\  
& Ticketing                                                                     & BugTrace (in-house tool)
 \\  \midrule
\multirow{8}{*}{\begin{tabular}[l]{@{}l@{}}Infrastructure\\mgmt\end{tabular}} &  CI/CD                       & Jenkins, KDM + Ansible                                                                      \\
& Repo Management                                                        & \begin{tabular}[l]{@{}l@{}}JFrog Artifactory (internal),\\ Sonatype Nexus (customers)\end{tabular} 
                            \\
& Config/Provisioning                                                          & KDM                                                                                               \\
& Release Management                                                           & KDM + custom bash scripts                                                                          \\
 & Containerization                                                        & Docker                                                                                            \\
& Monitoring                                                                   & \begin{tabular}[l]{@{}l@{}}KDM, Apache Kafka,\\ElasticSearch\end{tabular}
                            \\ 
& Logging                                                                      & Logstash                                                           
     \\  \midrule
Collaboration & Communication                                                  & \begin{tabular}[l]{@{}l@{}}Microsoft Exchange,\\Skype for Business, Slack\end{tabular}         
 \\ \bottomrule
\end{tabular}
\vspace{-6mm}
\end{table}

\subsubsection{Tooling}
The teams at \klop{} use a plethora of tools, as it can be observed in Table~\ref{tab:tooling}. The high number of adopted tools is a side effect of the coexistence of two solutions, the legacy desktop product and the new web application, which rely on different technology stacks.

The tools can be broadly grouped into three main categories, that is, \textit{development}, \textit{infrastructure management}, and \textit{collaboration}. Next, we comment on the use of only the tools that are relevant to the presentation of the case study.

In the development category, Jira plays a pivotal role in handling the backlogs of the Dev, Architecture, QA teams, each with slightly different modalities. The Dev team relies on a \textsc{Scrum} board with user stories (broken down into sub-tasks) that transition into four states (\textit{To do}, \textit{In progress}, \textit{Fixed/Developed}, and \textit{Done}). The DevOps team uses a Kanban board with six states (\textit{To do}, \textit{Planned}, \textit{In progress}, \textit{Done}, \textit{QA}, and \textit{Released}). Within the board, the team graphically groups tasks together to separate the high-priority one from those related to, for example, infrastructure maintenance.  Finally, the Architecture team  uses \textit{Epics} to group related user stories but does not rely on any specific type of board.

Another important development tool is Confluence, a workspace that stores all the knowledge base of the Klopotek ecosystem, from QA reports and deliverables to FAQs and how-tos.

Among collaboration tools, \klop{} rely on Microsoft Exchange for email and calendar. Email is the primary means of communication, used mostly for asynchronous communication with the Berlin site, but also for more formal internal communication at \klop{}. Microsoft Exchange is also used because it integrates with Skype for Business, which is the preferred instrument for video- and conference-calls with Berlin, albeit it is sometimes used for instant messaging (IM) too. 

Other than Skype, Klopotek has adopted Slack as the means of site-specific, technical-oriented communication. In fact, there are two separate workspaces, one for each of the two development sites. Regarding \klop{}, Slack is mostly used for quick questions, code snippet and file sharing, and one to one communication. Channels are instead used to gather information from external sources. At the time of the case study, the following integrations were already active: GitLab, Jenkins, Eclipse Code Sharing, and Google Drive.

\subsection{Data Collection}\label{sec:data}

Overall, the case study lasted for about four months, from December 2018 to April 2019.
The collection of data happened along the entire duration of the case study. There were four main data sources, that is, \textit{documentation}, \textit{direct observations}, \textit{tools}, and \textit{interviews}, as detailed next. 


\subsubsection{Documentation}

During the case study, we had access to documentation of different kind, such as PowerPoint presentations (describing the company's development process at a higher level, organization, and product architecture) and Excel spreadsheets (detailing project planning and performance analysis). Other helpful sources of documentation were the \textit{Strengths, Weaknesses, Opportunities, Threats} (SWOT) \cite{Helms2010-swot} and \textit{Balanced Scorecard} \cite{Lee2000-bsc} analysis reports, which helped us understand the areas perceived by developers and management as susceptible to improvement.

Overall, the pieces of information retrieved from documentation sources were used to triangulate, integrate, and confirm those obtained through the direct observation and semi-structured interviews, as described next.

\subsubsection{Direct observations} 

Regarding direct observations,  we could attend 18 different sessions (see Table~\ref{tab:observations}). The company allowed us to participate in one Grooming session, seven Daily Meetings, and two Sprint Reviews with the development teams. We agreed  not to take part in Retrospective meetings to avoid the risk of limiting the free expression of problems reported by the development team members. We also had the chance to participate in six \klop{} Management Board meetings and two Technical Board meetings. 

\begin{table}[t]
  \setlength{\abovecaptionskip}{-1pt}
  \caption{List of the 18 formal sessions of direct observation.}
  \label{tab:observations}
  \footnotesize
    \begin{tabular}{ll}
        \toprule
        \textbf{Date} & \textbf{Session} \\
        \midrule
        05 Dec.  2018 &  Management Board Meeting \\
        10 Dec.  2018 &  Management Board Meeting  \\
        11 Dec.  2018 &  Management Board Meeting \\
        07 Jan.  2019 &  Grooming \\
        08 Jan.  2019 &  Daily Meeting \\
        09 Jan.  2019 &  Daily Meeting \\
        09 Jan.  2019 &  Sprint Review \\
        10 Jan.  2019 &  Daily Meeting \\
        14 Jan.  2019 &  Daily Meeting \\
        15 Jan.  2019 &  Daily Meeting \\
        17 Jan.  2019 &  Daily Meeting \\
        18 Jan.  2019 &  Daily Meeting \\
        23 Jan.  2019 &  Sprint Review \\
        15 Jan.  2019 &  Technical Board Meeting \\
        24 Jan.  2019 &  Technical Board Meeting \\
        31 Jan.  2019 &  Management Board Meeting \\
        04 Mar.  2019 &  Management Board Meeting \\
        25 Mar.  2019 &  Management Board Meeting \\
        \bottomrule   
    \end{tabular}
    \vspace{-5mm}
\end{table}

The first session that we could observe was the \klop{} Management Board Meeting held in Dec. 2018, when a first draft of the SWOT analysis report for the year 2018 was discussed. This opportunity turned out to be fundamental for the case study because it allowed us to get an immediate understanding of the internal problems and goals, and thus inform the rest of the observations. 

Finally, we found it essential that Management allowed us to have one desk in the open-space office shared by all development teams, to directly observe how collaboration and communication happened daily.

\subsubsection{Tools}
Beside the desk in the open-space office, the managers allowed us to have credentials to access most of the tools used by the teams. We  focused on Confluence, Jira, and Slack. Having access to the Confluence knowledge base was helpful in retrieving Sprint Review and Retrospective reports as well as obtain detailed information about the architecture and development of the products, testing, and CI/CD processes.  Regarding Jira and Slack, the former was useful for understandings the workflow of the development teams (e.g., task assignments), whereas the latter was used to complement the direct observations and have access to the interactions that did not happen  (e.g., chats). 

\subsubsection{Interviews}

Along the case study, three sessions of semi-structured interviews\footnote{The scripts used in each interview session are available at https://figshare.com/articles/ICGSE\_2020\_Interviews\_material/11481759} were conducted: \textit{initial} (exploratory), \textit{intermediate} (preparation), and \textit{follow-up} (feedback).  

Initial, exploratory interviews were conducted at the beginning of our case study, around mid-December, after participating in the first three \klop{} Management Board meetings (see Table~\ref{tab:observations}). Since the intention was to clarify the results of the SWOT analysis, these interviews involved all the seven members of the Management Board. The interviews lasted about 15-20 minutes. 

Intermediate interviews were conducted in late January. 
The interviews lasted about 15-20 minutes each and also involved  members of development  teams (Dev, QA, Core, and DevOps), for a total of twenty-two participants. In this case, the template for guiding the interviews was structured in three parts. The first part gauged the experience of the interviewees; the second one focused on how they used tools, in particular, Outlook, Slack, Jira, and Confluence; the third one contained different questions depending on the role of the interviewee, aimed at uncovering job-specific themes.

The follow-up interview sessions were conducted to collect post-change feedback. As detailed in the next Sect.~\ref{sec:changes}, we introduced two changes, and feedback collection happened in mid-February and at the end of March. The goal here was to elicit feedback (e.g., effectiveness, appreciation) of the change introduced. After the first change, we interviewed Dev and QA team members (13 in total), whereas after the second, we also interviewed Core and DevOps team members (20 in total). Each interview lasted about 10 minutes and contained both closed- and open-ended questions.

\subsection{Changes}\label{sec:changes}

\begin{figure}[tb]
  \centering
  \includegraphics[width=\linewidth]{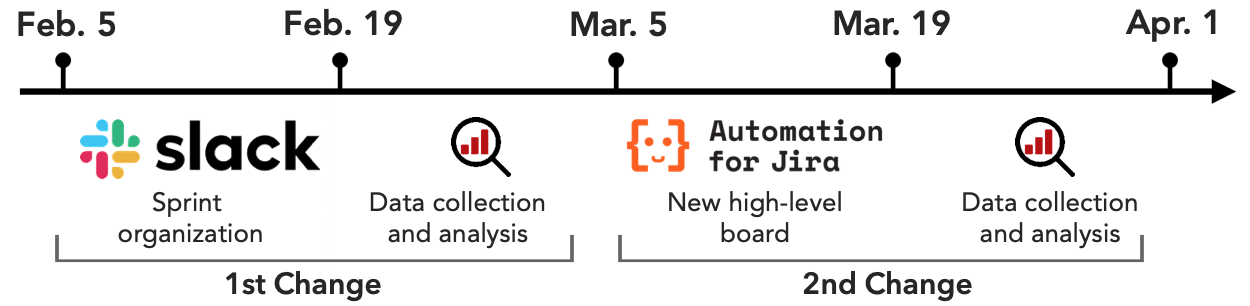}
  \setlength{\abovecaptionskip}{-3mm} 
  \setlength{\belowcaptionskip}{-8mm} 
  \caption{Timeline of the two introduced changes.}
  \label{fig:timeline}
\end{figure}

In Fig.~\ref{fig:timeline} it is illustrated the timeline of the two changes introduced at \klop{}. After initial observations and discussions with management, we learned that they  (i) were seeking solutions to streamline communication, both inter- and intra-site, and (ii) wanted to make the best of Jira, which was chosen as the core of their collaboration infrastructure. For each change, we observed its effects along an entire Sprint iteration (i.e., two weeks), at the end of which feedback was collected using semi-structured interviews, to understand whether the change was to be accepted or rejected and if it needed adjustments.

The \textit{first change} (Feb. 5 -- Feb. 19) focused on communication. Its goal was to promote more rational use of the communication tools, in particular to foster the use of Slack for internal communication instead of email and Skype for Business, which instead had to be used for all the external communication  (e.g., for messages and calls between \klop{} and Berlin or customers). In particular, three new public channels were added, \texttt{\#dev}, \texttt{\#qa}, and \texttt{\#sprints}. 

The expected benefit of the proposed change was making the internal conversations searchable and organized in proper channels, while also avoiding information to be fragmented over multiple channels and reducing the clutter of email inbox.

Regarding the execution, at the beginning of the selected Sprint iteration, a meeting was arranged with the members of the development teams involved in the change, i.e., Dev and QA. Here, the change was explained, and instructions were given, such as avoid using direct private messages in Slack for team-relevant communication. Eventually, we concluded the meeting with a Q\&A session to resolve any doubts. 

Other than the participant feedback collected through follow-up interviews, we retrieved communication logs from Slack to analyze both quantitatively and qualitatively the effects of the proposed change.

The \textit{second change} (Mar. 5 -- Mar 19) focused on the Jira tool and had a two-fold goal. First, empowering managers through the creation of a high-level board in Jira---to get a glimpse of the status of user stories---separated from the detailed board with sub-tasks assigned to developers and testers. Also, two new columns relevant for the QA team were added to the new board for stories ready to be tested and those currently under test, thus generating the following state transitions: \textit{To do}, \textit{In progress}, \textit{Read to test}, \textit{Testing}, and \textit{Done}. 
Second, through the integration of a Jira plugin,\footnote{https://automationforjira.com}, we introduced automation rules to  automatically map the states between low-level tasks in the detailed board and high-level user stories in the Management board. Also, the plugin allowed the notification of Jira events via Slack rather than email, for better integration between the two tools.

The expected benefits of the proposed changes were relieving users from the tedious duties of managing task states manually, increasing the visibility of important events via Slack, and reduce email overload.

Regarding the execution, at the beginning of the selected Sprint iteration, we arranged another preparation meeting, this time also involving the Management and Technical Board members, other than the other development teams. Here we illustrated the intended use of the new Management board as well as the automation rules offered by the Jira plugin and how to define custom ones.

Also, for the second change, we collected Slack communication logs along with follow-up interviews.

\section{Findings}\label{sec:findings}

In this section, we first report the findings from the direct observation and the initial/intermediate interviews (Sect.~\ref{sec:findings-obs+interviews}). Then, we report the findings from the two changes (Sect.~\ref{sec:findings-exp}).

\subsection{Direct Observation and Initial/Intermediate Interviews}\label{sec:findings-obs+interviews}

We conducted semi-structured interview sessions with twenty participants, in particular five developers (D1-D5), two testers (QA1-QA2), three members of the DevOps team (DO1-DO3), two members of the Architecture team (A1-A2), the four team leaders (one from each of the development teams, TL1-TL4), the and four managers (M1-M4). 

Four participants are female (19\%). Regarding the working experience, at the time of the case study, the managers and team leaders had been working at \klop{} for over three years, and had an overall working experience of 10+ years. The developers were less experienced: six had been working at \klop{} for less than a year and had an overall experience ranging between 1 and 5 years; the other six developers were more senior, with an overall working experience between 5 and 10 years, of which 1 to 3 years spent at \klop{}. 


\subsubsection{Communication Issues}
One of our goals for the case study was to understand how communication tools were used. We directly observed that most of the internal communication at \klop{} happened , especially within each of the development teams, whereas email was the primary means of computer-mediated communication.
During the preliminary interview sessions, we asked the participants to quantify the average number of `internal' emails sent daily and their main purposes. Seven participants (35\%) reported sending 10 or more internal emails per day, six (30\%) between 5 and 10, and the other seven respondents (35\%) less than 5. 
Regarding the main purposes for sending internal emails, the most reported reason (13 mentions) was the notification of events (e.g., user story masked completed on Jira, updated a documentation page on Confluence. 
The other two most cited  reasons  were \textit{ex aequo} (8) information requests and responses and event organization (e.g., scheduling calls or meetings).


After email usage, we focused on understanding how Slack was being used at \klop{}.  We chose to restrict the analysis of Slack usage statistics to the time-window of Oct. 2018-Jan. 2019 because before then, Slack configuration was different compared to when we started the case study (each project used to have a dedicated channel), but that setup had been already abandoned because deemed dispersive. As such, at the beginning of our case study, we analyzed what public and private channels were available and who had subscribed to them (Table~\ref{tab:channels-subs}).
Please, note that we received n=18 answers this time and that the channel mappings for managers M3 and M4 were not collected since they acknowledged not having really used Slack yet at that time. 

Regarding the available channels, \texttt{\#core}, \texttt{\#devops}, and \texttt{\#kdm} were used for receiving Jenkins notifications  from, respectively, the \textsc{Core} back-end framework component, the tests executed by the DevOps team, and the deployment of new KDM features. The channels \texttt{\#components-ui} and \texttt{\#components-ui-commits} were used to collect notifications from  GitLab regarding the framework front-end component.
The channels \texttt{\#core-internal}, \texttt{\#dev-internal}, and \texttt{\#devops-internal} were the private channels used by the respective team for intra-team communication. 

As for the channel subscriptions, the results in Table~\ref{tab:channels-subs} show that QA members and Managers had subscribed to very few channels. Instead, the DevOps, Core, and Dev team members (in order) had subscribed to more channels, mostly private (shown in grey).
All the eighteen respondents reported using the Slack desktop or web app as their daily driver, with only five reporting having the app also installed on their phone.
Ten of them (\textasciitilde56\%) reported having set the notification preference as  `\textit{Direct messages, mentions \& keywords},' with the others equally distributed between  `\textit{All new messages}' (22\%) and `\textit{Nothing} (22\%).'

\begin{table*}[ht]
\setlength{\abovecaptionskip}{-1pt}
\caption{Observed mapping of channel subscriptions per interview respondent (n=18). Channels not related to development activity (e.g., \texttt{\#random}, \texttt{\#general}) are omitted. Rows in grey indicate private channels.}
\label{tab:channels-subs}
\footnotesize
\begin{tabular}{l|cccccc|ccc|cccc|ccc|cc}
\toprule
                                    & \multicolumn{6}{c|}{\textbf{Dev}} & \multicolumn{3}{c|}{\textbf{QA}} & \multicolumn{4}{c|}{\textbf{DevOps}} & \multicolumn{3}{c|}{\textbf{Architecture}} & \multicolumn{2}{c}{\textbf{Management}} \\
\multirow{-2}{*}{\textbf{Channels}} & D1  & D2  & D3  & D4  & D5  & TL1 & QA1    & QA2    & TL2   & DO1     & DO2     & DO3     & TL3    & A1        & A2        & TL4        & M1                  & M2                 \\ 
\midrule
\#components-ui                     &     &     &     &     & x   & x   &        &        &       &         &         &         &        &           & x         &            &                     &                    \\
\#components-ui-commits             &     &     &     &     &     & x   &        &        &       &         &         &         &        &           &           &            &                     &                    \\
\#core                              &     & x   &     & x   &     &     &        &        &       & x       &         &         & x      & x         & x         & x          & x                   &                    \\
\#devops                            &     & x   &     &     &     &     &        & x      &       & x       & x       & x       & x      &           & x         & x          & x                   &                    \\
\#kdm                               &     &     &     &     &     &     &        &        &       & x       & x       & x       & x      & x         &           &            &                     &                    \\
\rowcolor{gray}
\#core-internal                     &     &     &     &     &     &     &        &        &       & x       & x       & x       & x      & x         & x         & x          &                     &                    \\
\rowcolor{gray}
\#devops-internal                   &     &     &     &     &     &     &        &        &       & x       & x       & x       & x      &           &           &            &                     &                    \\
\rowcolor{gray}
\#dev-internal                      & x   & x   & x   & x   & x   & x   &        &        &       & x       & x       & x       & x      & x         & x         & x          &                     &                    \\
\rowcolor{gray}
\#techboard                         & x   & x   &     &     &     &     &        &        &       & x       &         &         &        &           & x         &            &                     &                  \\
\bottomrule
\end{tabular}
\end{table*}

Regarding the integrations, we found that Slack had been integrated with Jira and GitLab to receive event notifications. During the preliminary and intermediate interviews, we asked about the perceived usefulness of these integrations. Only one participant reported using  both integrations, whereas the others had activated either Jenkins or GitLab. They found the stream of notifications from both tools to be overkill because the same notifications would arrive via email.

\begin{figure}[t]
	\centering
	\includegraphics[width=\linewidth]{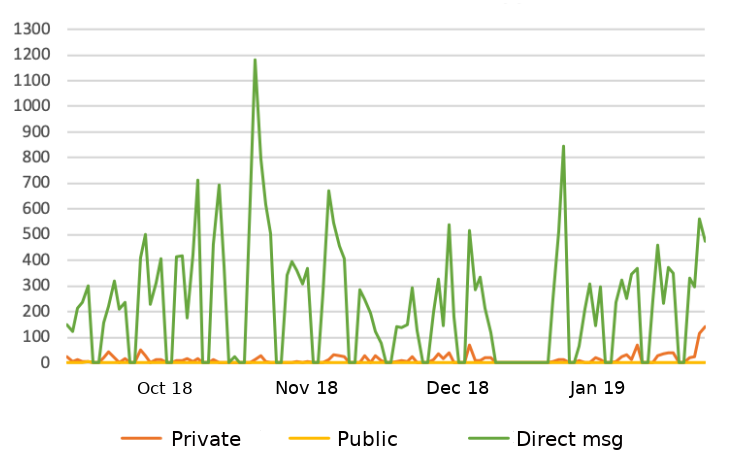}
	\setlength{\abovecaptionskip}{-8pt} 
	\setlength{\belowcaptionskip}{-4mm} 
	\caption{The observed distribution of messages sent over Slack by type (Oct. 2018 to Jan. 2019).}
	\label{fig:slack_msg_type}
\end{figure}

Next, we analyzed whether and how Slack  channels were used. Fig. \ref{fig:slack_msg_type} shows the distribution of messages sent to public channels, private channels, and as direct messages. We notice that, after abandoning the previous configuration, Slack was being used almost exclusively as an IM tool for sharing direct text messages. Otherwise, communication happened via private channels and almost never publicly.

Finally, albeit without access to logs, we directly observed that Skype for Business was also used for text-based, one-to-one chats.

\begin{tcolorbox}[standard jigsaw,
	title=Takeaways: Communication-related issues,
	opacityback=0]
	\small
	Direct observation and interviews revealed that internal communication at \klop{} was `unstructured:'
	\begin{itemize}[leftmargin=\parindent,align=left,labelwidth=\parindent,labelsep=3pt]
		\item Information overload exacerbated by the use of email for internal communication (notifications, info requests \& responses, event scheduling).
		\item Slack  used mostly for internal communication via direct messaging or private channels.
		\item Unclear rules about which medium to choose between email, Slack, and Skype for Business.
	\end{itemize}
\end{tcolorbox}\vspace{-2mm}

\subsubsection{Workflow Issues}
During the interviews, we asked about the \textsc{Scrum} workflow adopted at \klop{}. Albeit respondents acknowledged some of the expected and well-known benefits of the Agile methodology (e.g., time-boxed development, rapid client feedback, stand-up meetings, and Sprint retrospectives), there were also some complaints about it as well as related to the extent of tool support. 

First, developers and testers complained about the uneven workload that tends to spike at the end of each Sprint release; albeit they acknowledged this as a condition intrinsic to time-boxing, they also thought the problem was partially due to a planning issue, specifically, ``\textit{the overly optimistic cost estimation of the user stories in the backlog}'' (QA1).

Other issues were mentioned that make the workflow ``\textit{not perfect}'' (A1). DevOps team members felt that ``\textit{the general workflow does not involve us to the right extent}'' (DO1) and that ``\textit{this often causes misalignment with the other teams, especially with the Dev team}'' (DO2). 

Second, regarding tool support, while Jira was generally appreciated, there were also complaints about it being confusing at times. After seeking clarification, we found out that the managers and team leaders thought that  all the details available in the Jira board were sometimes distracting, for example, when they just want to quickly assess the advancement status of a Sprint, like, which user stories are completed, under test, and to do. Still, they acknowledged that those details are necessary to the development teams nonetheless.

The developers had complaints about Jira too. In particular, they generally reported the need for manually executing actions to synchronize the status of each user story after changing those of its sub-tasks.

\begin{tcolorbox}[standard jigsaw,
    title=Takeaways: Workflow-related issues,
    opacityback=0]
    \small
Direct observation and interviews revealed the following problems related to the collaboration workflow in Jira:
\begin{itemize}[leftmargin=\parindent,align=left,labelwidth=\parindent,labelsep=3pt]
    \item \textsc{Scrum} board filled with details useful for the developers but `distracting' for managers and team leaders.  
    \item Several, tedious actions needed in the \textsc{Scrum} board for maintaining up-to-date and in sync the status of issues and their sub-tasks.
\end{itemize}
\end{tcolorbox}\vspace{-2mm}

\subsection{Changes}\label{sec:findings-exp}

Building upon the observations and interview outcomes, we implemented the two changes detailed earlier in Sect. \ref{sec:changes}. Here we report our findings.

\subsubsection{First Change: Reinforcing the Use of Slack}

The focus of the first change was on reinforcing the use of Slack as a central hub for internal communication. To counteract the communication issues that emerged from the interviews and direct observation,  during the change, Slack became the tool of choice for channeling all the internal communication. At the same time, the use of Skype for Business and Outlook was recommended for communicating (i.e., email, voice call, screen sharing) with remote sites and customers.

To foster the use of Slack for internal communication,  we added to the \klop{} workspace three new channels with the following usage rules that aimed at replacing the use of internal emails for development, testing and Sprint organization entirely:
\begin{itemize}
	\item \texttt{\#dev} --- a public channel used by the Dev team to report problems, seek advice from others, and share code snippets; this channel replaced the private one \texttt{\#dev-internal};
	\item \texttt{\#qa} --- a public channel for hosting internal discussion of the QA team about quality and to notify user story completion;
	\item \texttt{\#sprints} --- a public channel for sharing Sprint retrospective reports and urgent communications about the current Sprint, which need immediate managers' attention. 
\end{itemize}   

As a result, by using Slack, all the Sprint-related messages were collected in one public place accessible by everyone at \klop{}, while also helping to reduce email overload. Please, note that, other than the new three channels, the Slack configuration during the execution of first change also included the previously existing channels reported in Table~\ref{tab:channels-subs}, which we did not analyze.

Finally, to ensure that all the participants used the tool as intended during the change, we added to Confluence a new documentation page about Slack, with channel descriptions, instructions, and advice for a proper use such as avoid using direct messages and creating new private channels for communication relevant for a whole team.

\textbf{Quantitative Analysis Results}. As already mentioned in Sect. \ref{sec:changes}, data collection for the first change lasted for two weeks (i.e., one entire Sprint) from Feb. 5 to Feb. 19. After collecting the logs of messages exchanged in Slack, we performed both quantitative and qualitative analysis by first counting the number of active users, messages sent, and mentions per user in the new channels; then, we performed the content analysis of the messages. 

During the Sprint, 125 messages were sent in the three new channels, distributed as follows: 80 messages in \texttt{\#qa}, 32 in \texttt{\#sprints}, and 13 in \texttt{\#dev}. The \texttt{\#qa} channel was also the most active in terms of number of active participants\footnote{An active participant is a team member who shared at least one message.} (9) and shared \texttt{@username} mentions (101), followed by \texttt{\#sprints} (6 active users and 12 mentions) and \texttt{\#dev} (5 and 2).

To understand what kind of information was exchanged in the channels during the Sprint, we manually performed the content analysis of the messages using the coding schema designed by \citeauthor{Stray2019}~\cite{Stray2019} in a similar study. The coding schema is reported for convenience  in Table~\ref{tab:coding}.

\begin{table}[t]
  \setlength{\abovecaptionskip}{-1pt}
  \caption{The Coding schema used in this study (from \cite{Stray2019}).}
  \footnotesize
  \label{tab:coding}
    \begin{tabular}{lll}
        \toprule
        \textbf{Category} & \textbf{Definition} & \textbf{Example} \\
        \midrule
		\begin{tabular}[l]{@{}l@{}}Coordination / \\General info\end{tabular}  & \begin{tabular}[l]{@{}l@{}}Giving general\\information\end{tabular} & \begin{tabular}[l]{@{}l@{}}\textit{Today I’m working from}\\\textit{home and won't take part}\\\textit{in the scrum daily meeting}\end{tabular} \\
		\hline
		\begin{tabular}[l]{@{}l@{}}General\\discussion\end{tabular} & \begin{tabular}[l]{@{}l@{}}Q\&A regarding\\general topic\end{tabular}  & \textit{Is this part of the current Sprint?} \\
		\hline
		\begin{tabular}[l]{@{}l@{}}Problem-focused\\communication\end{tabular} & \begin{tabular}[l]{@{}l@{}}Technical questions\\and discussions of\\possible solutions\end{tabular} & \begin{tabular}[l]{@{}l@{}}\textit{Is there anything that we can}\\\textit{do to align the two frames?}\end{tabular} \\
		\hline
		\begin{tabular}[l]{@{}l@{}}Technical\\information\end{tabular} & \begin{tabular}[l]{@{}l@{}}Giving technical\\information\end{tabular} & \textit{STREAMRSCM-646 closed} \\
		\hline
		Socializing & \begin{tabular}[l]{@{}l@{}}Messages used\\for socializing\end{tabular} & \begin{tabular}[l]{@{}l@{}}\textit{[If you want me to}\\\textit{close this task] 10€}\end{tabular} \\
		\hline
		Emoji & \begin{tabular}[l]{@{}l@{}}Emojis sent\\by users\end{tabular} & \textit{:pray:}, \textit{:smile:} \\
        \bottomrule   
    \end{tabular}
    \vspace{-3mm}
\end{table}

The coding was performed iteratively by the first two authors. In the first iteration, we randomly selected a sample of 40 thematic units. Each thematic unit \cite{Calefato2012-phd} was obtained by identifying in the log a set of messages sent by one user, which have the same intent (i.e., are conceptually linked together) even if split into multiple, non-consecutive utterances. Otherwise, had we chosen a few random messages and ignored their `context,' the coding procedure would have been impossible. Furthermore, we selected a number of thematic units from each of the three channels proportional to the overall number of messages sharing therein during the experiment. 

The first coding iteration was performed individually, with an inter-rater agreement of 85\%; then, the two coders participated in a  meeting to resolve disagreements. The second iteration followed a similar process on a new set of 40 thematic units. Because the measured inter-rater agreement was high (90\%), after resolving the disagreements, we decided that the remaining units could be safely coded by one coder only. (For simplicity, from now on, we will use the term message.)

\begin{figure}[t]
  \includegraphics[width=\linewidth]{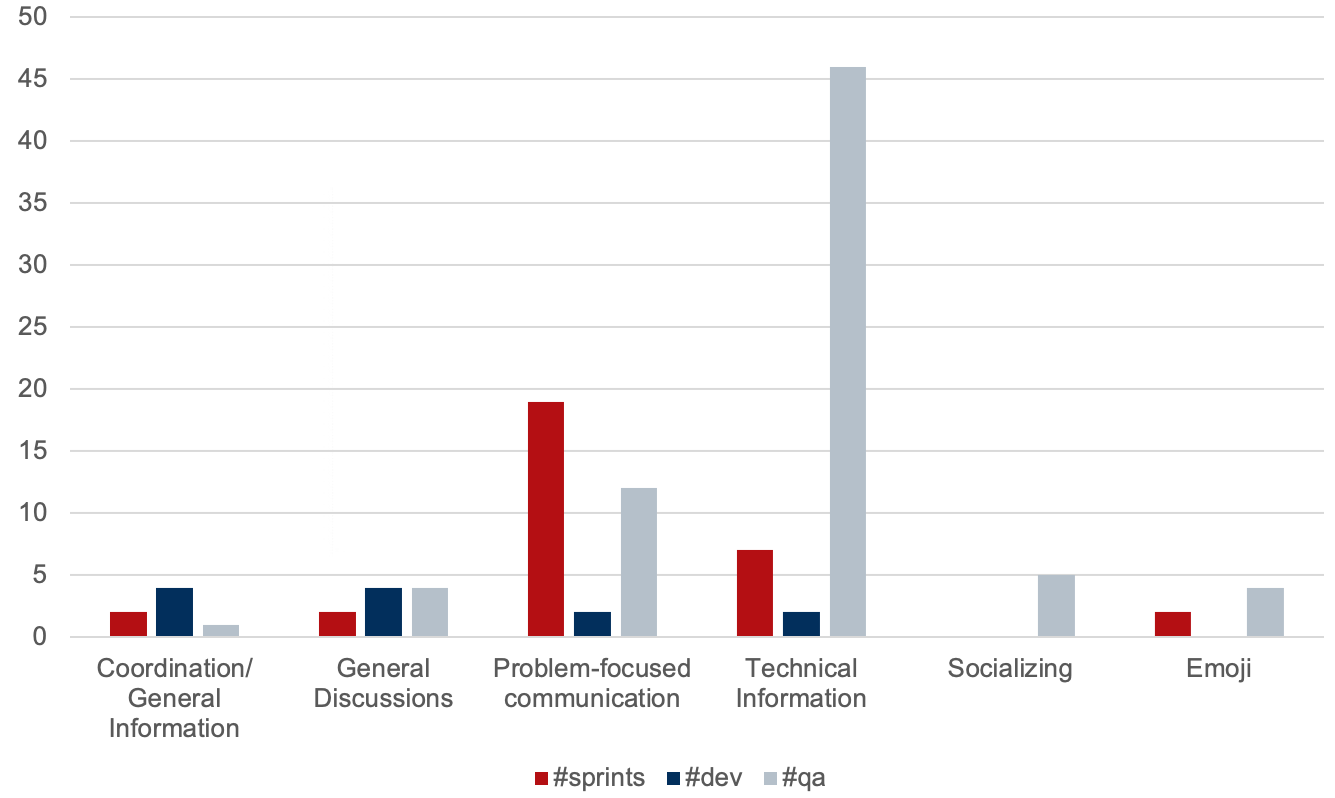}
  \setlength{\abovecaptionskip}{-8pt} 
  \setlength{\belowcaptionskip}{-6mm} 
  \caption{The results of content analysis with the distribution of categories for the new channels (Sprint Feb. 5-19).}
  \label{fig:coding_res}
\end{figure}

Fig.~\ref{fig:coding_res} shows the distribution of coded messages across the three new channels. The most common intents for communicating were sharing \textit{Technical Information} (55 messages) and  resolving issues (\textit{Problem-focused Communication},  33 messages). A very few messages were exchanged for general purpose Q\&A  (\textit{General Discussions}, 10), to coordinate (\textit{Coordination/General Information}, 7), and socialize (\textit{Emoji} 6, \textit{Socializing} 5).

Furthermore, Fig.~\ref{fig:coding_res}  shows that, overall, the \texttt{\#qa} channel was the most active one during the experiment. Also, contrary to expectations, we found  a higher percentage of \textit{Technical Information} messages shared in the  \texttt{\#sprints} channel as compared to \texttt{\#dev}, mostly for matters related to releases and execution environment configurations. Albeit communication of this kind is important for Sprint execution (e.g., feature demonstration, Sprint review), this result shows an improper use of the \texttt{\#sprints} channel since it was conceived for sharing notification about urgent, blocking issue---to make it more focused, with messages about the previous Sprints always at hand---whereas, technical-related content had to be shared only in the other two channels. 
Relatedly, despite the small difference, \klop{} team members shared more messages coded as \textit{Coordination/General Information} in \texttt{\#dev} rather than in \texttt{\#sprints}, a further proof of improper use of the latter channel. 

Finally, regarding the few messages coded as \textit{Socializing} and \textit{Emoji}, we note that the use of Slack at \klop{} is intended exclusively for work-related matters. Therefore, it is likely that socialization happens through other channels (e.g., ) not considered in this case study.

\textbf{Qualitative Analysis Results}. At the end of the Sprint, we conducted semi-structured interviews with 13 participants (the entire Dev and QA teams, TL3, TL4, M1, and M2) to get feedback about how the use of Slack affected internal communication at \klop{}.
First, we asked whether they happened to send emails and, if so, why. We found out that only 4 out of 13 fully complied with the instructions and used Slack instead of using of `internal' emails. 
The reasons that they mentioned included: (i) \textit{privacy concerns}, e.g., one developer did not want to negotiate time-off with a manager on a public channel, so they chose email instead of private messaging via Slack; (ii) \textit{notification overload}, i.e., one developer reported a bug found in the \textsc{Core} framework via email instead of using  \texttt{\#core} as established because the channel became overloaded by the continuous flow of Jenkins notifications, which ended up  making any human interaction impossible; (iii) \textit{IM replacement}, i.e., users had the (wrong) perception that Slack was better suited for short, instant messages and, hence, the use of emails was preferred for longer, more complex communication like discussing a feature from the current Sprint.

\begin{figure*}[tb]
    \begin{subfigure}{.48\textwidth}
        \centering
        \includegraphics[width=1.0\linewidth]{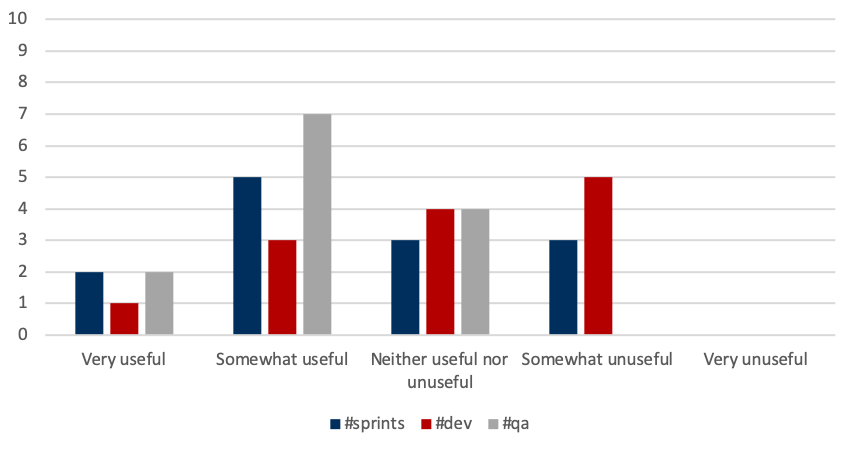} 
        \caption{First Sprint (Feb. 5-19, 2019)}
        \label{fig:a}
    \end{subfigure}
    \begin{subfigure}{.48\textwidth}
        \centering
        \includegraphics[width=1.0\linewidth]{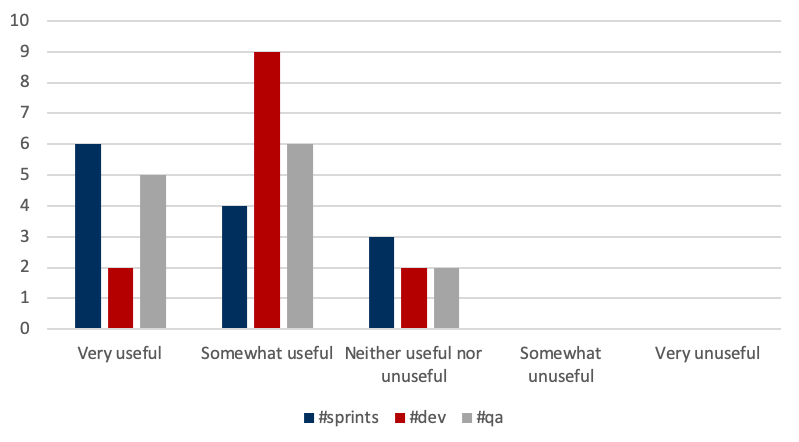} 
        \caption{Second Sprint (Mar. 5-19, 2019)}
        \label{fig:b}
    \end{subfigure}
    \setlength{\abovecaptionskip}{-1pt} 
    \setlength{\belowcaptionskip}{-8pt} 
    \caption{Perceived levels of usefulness of the three new channels (a) after the first Sprint and (b) after the second.}
    \label{fig:first_change_satisfaction}
\end{figure*}

Fig.~\ref{fig:a} reports the results of a survey for measuring the perceived level of usefulness for the three new channels (\texttt{\#dev}, \texttt{\#qa}, and \texttt{\#sprints}) on a 5-point Likert scale, anchored with the values 1=\textit{Very unuseful}, 2=\textit{Somewhat unuseful}, 3=\textit{Neither useful, nor unuseful}, 4=\textit{Somewhat useful}, and 5=\textit{Very useful}. Before commenting on the results, we point out that, during the follow-up interviews, two managers and the QA team leader acknowledged  that they always chose the mid-point category. They told us that they felt unable to fully evaluate the usefulness of the solution because still not used to Slack and also because some of these new channels were not a big part of their workflow.

Regarding the \texttt{\#dev} channel (median = 3), 5 out of 13 respondents reported finding it \textit{Somewhat unuseful}. When asked to elaborate during the follow-up interviews, they told us that because in the Dev team ``\textit{[they] are lucky to be co-located in the same open-space office}'' (D1),  is the preferred option for intra-team communication at \klop{}. Also, they seldom share a problem or ask for feedback team-wide (i.e., in the public channel); instead, they rather need communication tools to ``\textit{exchange direct messages during pair programming sessions}'' (D2). 

The \texttt{\#qa} channel was the most appreciated (median = 4), as 9 respondents found it \textit{Very} or \textit{Somewhat useful}. The main reason for the appreciation was that all the email notifications of completed user stories were replaced by messages in the public \texttt{\#qa} channels. QA team members found those emails to be ``\textit{dispersive, kind of wasted for a one-sentence-only message}'' (QA2), whereas with Slack notifications ``\textit{things are much faster and more immediate than emails}'' (D3). In general, the QA team felt that with Slack, their workflow had become ``\textit{simpler, a bit more streamlined}'' (QA2).

With respect to \texttt{\#sprints} (median = 4), the evaluation was slightly less positive than \texttt{\#qa}, with 7 respondents finding it \textit{Very} or \textit{Somewhat useful}, and 3  who thought it was \textit{Somewhat unuseful}. When interviewed, the participants said that they saw potential in it but also that the channel would ``\textit{take more than a Sprint to come to fully understand its benefits}'' and it would likely become ``\textit{more useful over time}'' (D1).
Therefore, we decided to collect survey responses from the same participants again, after another Sprint (i.e., two more weeks, see Fig.~\ref{fig:b}). Compared to the results from the previous Sprint, we notice that an increase in the level of perceived usefulness of the \texttt{\#dev} channel in the second Sprint as compared to the first one (median 4 and 3, respectively). No noticeable differences are present regarding the other two channels, instead.

\begin{figure}[tb]
    \centering
    \includegraphics[width=1.0\linewidth]{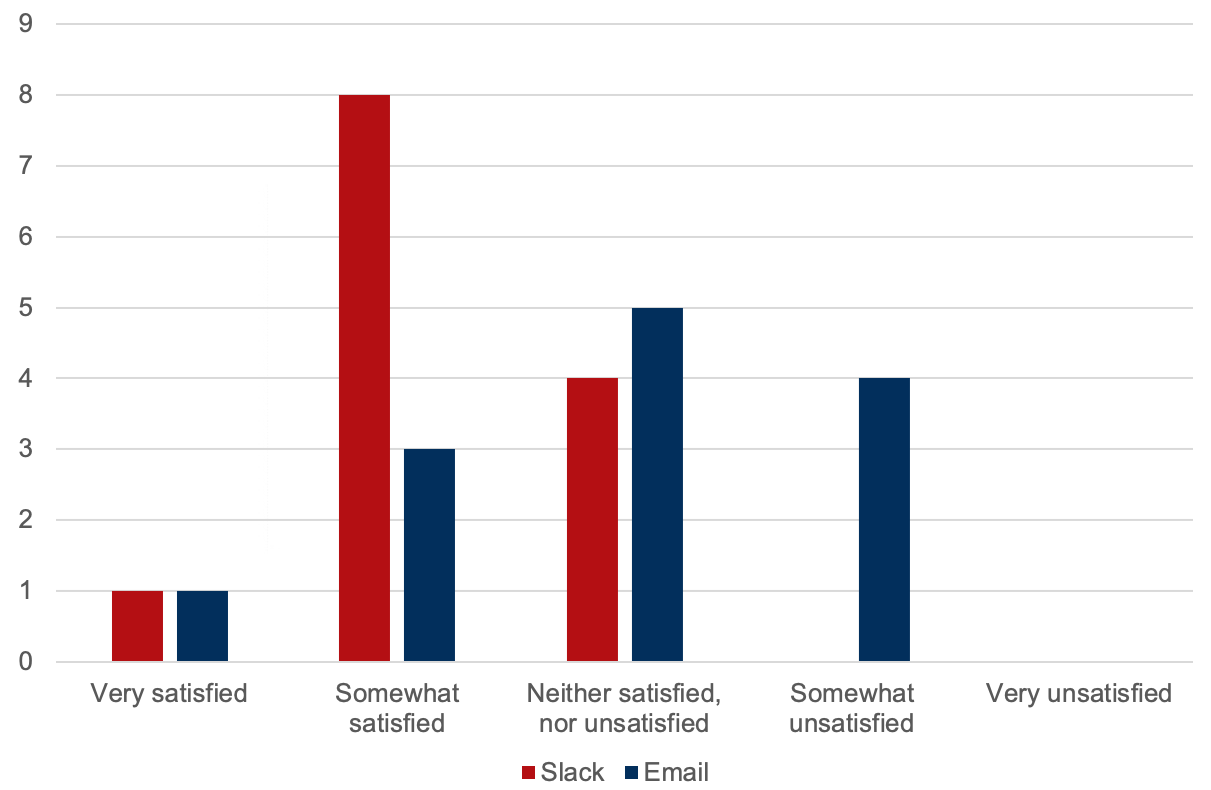}
    \setlength{\abovecaptionskip}{-1pt}
    \setlength{\belowcaptionskip}{-3mm} 
    \caption{Perceived levels of satisfaction: email vs. Slack.}
    \label{fig:slack_vs_email}
\end{figure}

Finally, during the follow-up interviews, we collected the perceived levels of satisfaction with using email internally as compared to Slack. Fig.~\ref{fig:slack_vs_email} shows that the  participants were more favorable to Slack (median = 4, \textit{Somewhat satisfied}) than email (median = 3, \textit{Neither satisfied, nor unsatisfied}), with no one feeling unsatisfied about the former.

\subsubsection{Second Change: Multiple \textsc{Scrum} Boards and Jira Automation Rules}

In the second change, we focused on improving the collaboration workflow in Jira. To counteract the workflow-related issues that emerged from direct observation and interviews, we first created a new, higher-level \textsc{Scrum} board meant in particular for the managers and team leaders. This new board shows only the main user stories, tasks, and bug reports, and provides at a glance a picture of how the current Sprint is going, with the original board showing the detailed sub-tasks useful for the Dev and QA teams.
Furthermore, the high-level board features two new intermediate states (columns): \textit{Ready to test}, for any issue not yet taken care of by QA, and \textit{Testing}, for issues currently under test. Therefore, during the second change, the issues in the new board transitioned into the following states: \textit{To do}, \textit{In progress}, \textit{Ready to test}, \textit{Testing}, and \textit{Done}.

The other proposed improvement was the installation of \textit{Automation for Jira}, a  third-party Slack plugin that allows the definition of automation rules in the form \textit{event}--\textit{condition}--\textit{action}. 
Examples of events are the creation and modification of issues; conditions can be expressed via JQL (Jira Query Language), regular expressions, and  graphically through the plugin; typical actions are sending emails and Slack messages through webhooks. 
For the sake of space, next, we report only a few examples of the ten automation rules defined for user stories while omitting those for bugs and tasks, which are nonetheless similar; we also point out that these rules have been activated for projects developed only at the Italian site and not shared with Berlin:
acmcopyright
\begin{itemize}
    \item \texttt{sprint\_started}: when a new Sprint is marked as \textit{Started} in Jira, send a notification to \texttt{\#sprints};
    \item \texttt{start\_implementation}: when the state of the first sub-task changes from \textit{Open} to \textit{Implementation}, move the parent user story from \textit{To do} to \textit{In progress}. 
    \item  \texttt{ready\_to\_test}: when all the sub-tasks are marked as \textit{Closed} (completed), move the parent user story into \textit{Ready to Test} and notify the testers in \texttt{\#qa};
    \item \texttt{all\_done}: when all the sub-tasks are closed and tested, the parent user story is marked as \textit{Closed} and a notification of user story completed is sent to \texttt{\#sprints};
    \item \texttt{version\_released}: when a new version of project in Jira is marked as \textit{Released}, send a notification to \texttt{\#sprints}.
\end{itemize}

The expected benefits of the second change were: 
(i) declutter the \textsc{Scrum} board for managers and team leaders and, by separating completed stories from those under testing,  allow them to identify possible bottlenecks if user stories pile up in the \textit{Ready to Test} column;
(ii) relieve the development teams from the burden of keeping Jira and the others up-to-date after every change.

\begin{figure}[t]
    \centering
    \includegraphics[width=1.0\linewidth]{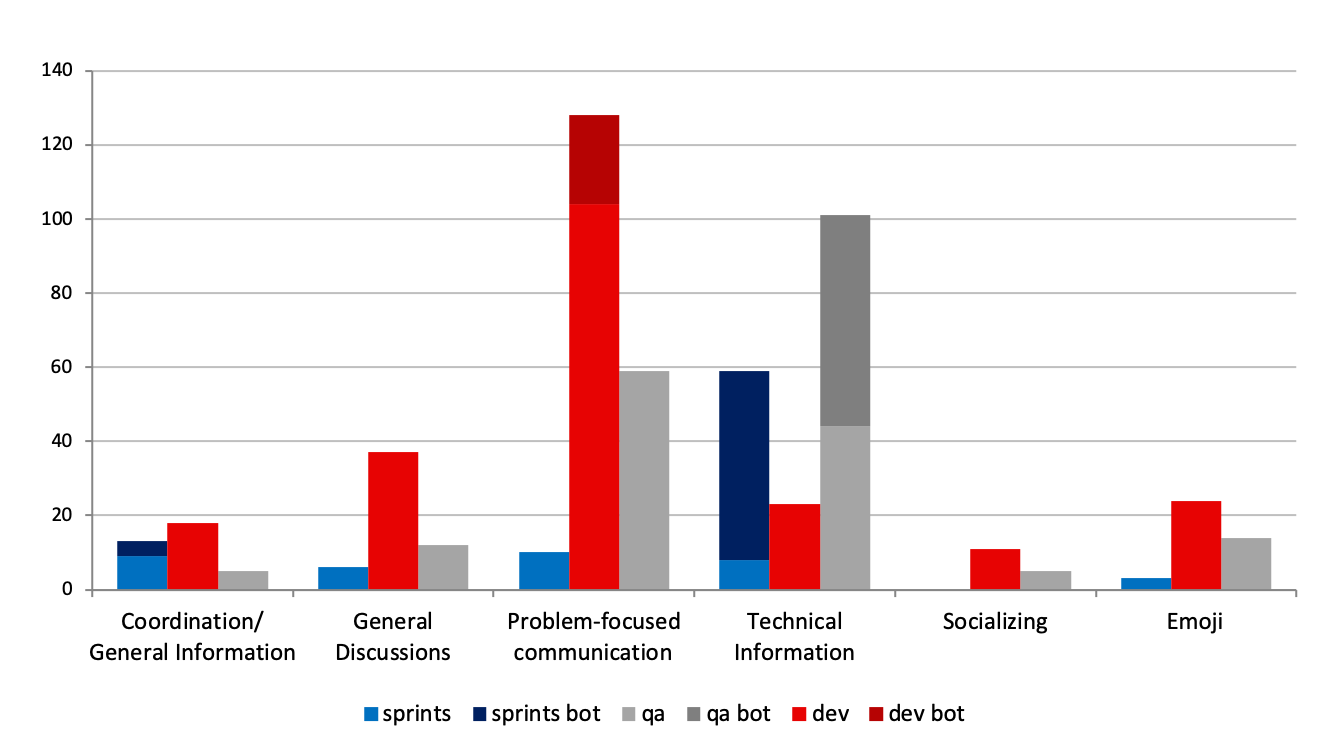} 
    \setlength{\abovecaptionskip}{-8pt} 
    \setlength{\belowcaptionskip}{-8mm} 
    \caption{Classification of messages exchanged between Feb. 19-Mar. 19 (plugin notifications shown on top of each bar, if any).}
    \label{fig:msgtypesS2}
\end{figure}

\textbf{Quantitative Analysis Results}. 
The data collection for the second change lasted for four weeks (two Sprints) from Feb. 19 to Mar. 19, during which 932 messages were published in the five active channels. Also in this case, we performed the content analysis of the messages using the same coding schema used before (see Fig.~\ref{fig:msgtypesS2}).

The \texttt{\#sprints} channel received 107 messages and 18 mentions from 16 active users. Compared to the first observation, there was reduction in  \textit{Problem-focused communication} (see Fig.~\ref{fig:coding_res} vs. Fig.~\ref{fig:msgtypesS2}); at the same time, there was an increase of the \textit{Technical Information} category, although most of the messages therein (86\%) are from the Jira automation plugin. The \texttt{\#dev} channel received 267 messages and 87 mentions from 17 users, with an overall in increase compared to the first observation. 
This increase was caused mainly by having part of the Dev team working remotely during these Sprints and also by the automatic notifications of defects found by QA (\textasciitilde 83\% of total messages).

Regarding the \texttt{\#qa} channel, it received 218 messages and 173 mentions from 15 users. Most of these messages (\textasciitilde56\%) are  notifications from the Jira automation plugin, such as `user story ready to test' or `bug resolved' with explicit mentions of some QA team members. The \texttt{\#core} channel was seldom used during the two Sprints (24 messages and 2 mentions from 8 users) and mostly for \textit{Problem-focused Communication}. The \texttt{\#devops} channel received 316 messages and 54 mention from 19 users. Most of the messages are coded as \textit{Problem-focused communication} whereas \textit{Technical Information} messages were mostly (60\%) automatic notifications from GitLab.

Finally, regarding the two most common categories, we found that, overall, most of the messages classified as \textit{Problem-focused Communication} are from users rather than plugin notifications (284 vs. 24) whereas the opposite is true for \textit{Technical information} (85 vs. 120). This difference is statistically significant (${\chi}^2=156.97$, $p<0.01$), meaning that people at \klop{} interact over Slack mostly for problem-solving. 

\textbf{Qualitative Analysis Results}. At the end of the Sprints, we interviewed all the twenty participants. From these follow-up interviews, we found that the high-level board was not used and appreciated by Management to the expected extent. Contrary to expectations, it was the Dev team to find it useful to get a ``\textit{bird's-eye view}'' (TL1) and an understanding of ``\textit{user stories' progress in the second half of a Sprint and know how many are still under development and to do, which is a critical piece of information to us}'' (D1). Instead, one manager (M3) admitted to having forgotten about it; still, he  added that he found the idea promising and intended to try it during the next Sprints. Another manager (M4) did not appreciate the lack of distinction ``\textit{between user stories for which development has not started yet  and those that went back into `To do' because of defects found}.'' He, therefore, suggested adding a new column \textit{Defect resolving} to handle the latter case. He also suggested adding a feature to quickly filter issues based, e.g, on the assignee, reporter, which would be useful during daily meetings and Sprint retrospectives.

Furthermore, we asked the participants about the perceived usefulness of the \textit{Automation for Jira} plugin. Four respondents found it \textit{Very useful} and seven \textit{Somewhat useful}, with only two managers finding it \textit{Neither useful, nor unuseful}. 
We also asked them to rank the ten automation rule in descending order of usefulness. The results are reported in Table~\ref{tab:rankrules} aggregated per team and overall. The top three ranked rules are \texttt{ready\_to\_test} (``\textit{[it] saves a lot of time not to have to notify testers when a new story is ready for them}'', D1), \texttt{start\_implementation} (``\textit{avoiding manual changes in Jira so convenient}'', D4), and (``\textit{interaction is more direct and immediate}'', QA1). While the ranks of the Dev and QA teams are very similar to each other, there are some differences with the managers, who are more interested not only in being notified about defects being found and fixed but also about the beginning and end of implementation. 

Thanks to the plugin, 85\% of the interviewees found their workflow to be improved, and the remaining found it unaffected by it. No participant found it to have a negative effect (``\textit{automation is always good}'', TL1), and all agreed to keep on using it afterward.

\begin{table}[t]
  \setlength{\abovecaptionskip}{-1mm}
  \caption{Perceived usefulness of automation rules (1=most useful, 10=least useful).}
  \footnotesize
  \label{tab:rankrules}
    \begin{tabular}{rllll}
        \toprule
        & \textbf{Dev} & \textbf{QA} & \textbf{Managers} & \textbf{Overall} \\
        \midrule
        1 & ready\_to\_test & ready\_to\_test & start\_implement. & ready\_to\_test \\
        2 & defect\_found  &  start\_implement. & defect\_resolved & defect\_found \\
        3 & defect\_resolved & defect\_resolved & defect\_found & start\_implement. \\
        4 & start\_implement. & sprint\_started & all\_done & defect\_resolved \\
        5 & sprint\_started & defect\_found & sprint\_started & sprint\_started \\
        6 & default\_subtasks & default\_subtasks & testing & default\_subtasks \\
        7 & testing & version\_released & ready\_to\_test & testing \\
        8 & version\_released & sprint\_completed & sprint\_completed & version\_released \\
        9 & sprint\_completed & testing & default\_subtasks & sprint\_completed \\
       10 & all\_done & all\_done & version\_released & all\_done \\
        \bottomrule   
    \end{tabular}
    \vspace{-6mm}
\end{table}

Finally, we asked the interviewees about the perceived usefulness of the channels \texttt{\#core} and \texttt{\#devops} that were not analyzed during the first change. Thanks to the feedback collected from the first Sprint, we applied some modifications in the policies of their use. First, we disabled the automatic notifications from Jenkins, which were deemed uninteresting \textit{per se} and were so numerous that they prevented any human interaction. Also, all participants were recommended to set their preferences to \textit{Direct messages, mentions \& keywords} not to miss any relevant notification.
Regarding \texttt{\#core}, most of the interviewees (8) found it \textit{Neither useful, nor unuseful}. Architecture team members A1 and TL4 agreed that this is because ``\textit{[we] rather chat  because architectural discussions are too long to type in}'' (A1).
The \texttt{\#devops} channel was better appreciated by the interviewees (4 \textit{Very useful} and 4 \textit{Somewhat useful}). 
Without Jenkins notifications plaguing it, the channel was found to be ``\textit{more focused, with only  relevant info}'' (DO1), as it allowed them to ``\textit{keep in touch with the Dev team and help them fix configuration issues}'' (DO2). Also, they appreciated that it is useful  ``\textit{not only to DevOps but everyone, being able to see when a request is made, it helps increase the awareness in the teams overall}''  (DO3).

\section{Discussion}\label{sec:discussion}

In this section, we first discuss the findings from our case study on how to improve tool support to collaboration at \klop{}. Then, we report the implications of our findings and discuss their limitations.

\subsection{Slack as a Central Hub of Communication}
After observing inconsistencies in the way communication tools were used at \klop{}, as a first change, we reinforced the use of Slack as the central hub for internal communication, while setting clear rules regarding other tools and channel usage. 

\subsubsection*{Slack fosters communication transparency}
Thanks to the use of public channels for sharing team-relevant pieces of information, the new Slack configuration introduced helped to make communication  at \klop{} more `transparent.'  As a further confirmation of this finding, we compared the Slack analytics from two months before and after the case study and noticed that the number of messages sent to public channels increased from 0.1\% to 4.4\% while, at the same time, direct messages  decreased from 92\% to 85\% and  those sent to private channels went down to 7.9\% from 10.6\%.
This increase in public messages was the result of both ensuring everyone complained with the recommended notification setting (i.e., no more messages went unnoticed) and managers getting more and more comfortable with Slack.

\subsubsection*{Slack is for everyone (to different extents)}
Overall, \klop{} teams liked using Slack as a central hub of communication and used it mostly  problem-solving. However, teams appreciated it to different extents. QA and DevOps teams were the most favorable, although it also grew on developers during the next Sprints, whereas the solution turned out to be not as suitable for the Architecture team members who found that their long, intra-team discussions and brainstorming sessions about the \textsc{Core} framework are better served .

\subsection{Jira Boards and Automation Rules} 
As a second change, we added to Jira a new high-level \textsc{Scrum} board while also introducing automation rules and notifications to Slack.

\subsubsection*{Managers need more time to adjust} 
The feedback interviews revealed that the high-level Jira board was not used by managers as expected---development teams seamed to appreciate it better---yet they told us they wanted to keep trying to use it. 
We observed similar behavior in the case of Slack. Therefore, our experience suggests that managers may need more time to adjust and get used to changes in the way collaboration tools support their workflow.

\subsubsection*{The more (automation) the merrier} 
Overall, the ten automation rules introduced in Jira were found useful, with 85\% of participants reporting a more streamlined workflow thanks to them. Besides, while automation rules may not be as useful for everyone, they are detrimental to no one. As such, \klop{} is considering expanding the set of automation rules also to other tools.

\subsection{Related Work} 

\subsubsection*{Beware of automatic notifications from development tools} \citeauthor{Calefato2016-hns}~\cite{Calefato2016-hns} proposed a Hub-and-Spoke model to loosely integrate software development and collaboration tools, and create a central hub to fight channel fragmentation and communication overload. They also implemented a prototype that used Slack as a central hub. In our case study, the first change restructured the existing Slack workspace configuration in a way that broadly follows the model recommended in \cite{Calefato2016-hns}. Our findings suggest that while the model helped to make  pieces information otherwise fragmented all available in one place, there is also a severe risk of  overload of automatic notifications from development tools connected to Slack, which can even hinder human interaction.

\vspace{-1mm}
\subsubsection*{Use separate, public channel for each team} \citeauthor{Stray2019}~\cite{Stray2019} investigated the use of Slack for coordination in a distributed software organization. While we focused on the intra-site communication of a global software company, we reused their coding schema to understand what was discussed on Slack at \klop{}. \citeauthor{Stray2019} observed mostly \textit{Problem-focused Communication} (48\%), followed by \textit{Technical Information} (20\%) and \textit{General Discussion} (15\%). If we ignore the automatic notifications, we consistently find that most of the messages shared during the case study belong to the \textit{Problem-focused Communication}.
Despite these differences in communication intent, we notice that \citeauthor{Stray2019} created a set of final guidelines that closely resembles the configuration of the Slack workspace in our case study, including (i) use public channels to foster open communication, (ii) create one channel per team, and (iii) aggregate and archive discussions in one place to make it searchable.

\subsection{Limitations}
As with every study, there are some limitations to our work. First, we followed a longitudinal, single-case design and, therefore, the general criticism of uniqueness applies. 
Another limitation related to generalizability is that we were allowed to analyze collaboration and communication only within the Italian site (quite limited in size), rather than between sites, where interaction also happened F2F. 
A second limitation relates to potential bias in our data collection. 
Some data was based on direct observations and semi-structured interviews conducted by one of the authors.
However, to reduce bias, multiple data sources were used to triangulate relevant pieces of information, and a protocol was agreed upon before conducting the interviews.
Finally, another limitation is that we did not analyze the effects of the implemented changes on communication happening through the other channels, such as F2F and email.

\section{Conclusions and Future Work}\label{sec:conclusions}

In this paper, we reported on a case study conducted at \klop{}, the Italian site of a large, distributed software company. 
We analyzed their Agile tooling and work environment, and proposed a couple of changes in the way they used Slack and Jira to improve the collaboration.
We found out that, overall, \klop{} accepted the restructured Slack workspace and the Jira automation rules, and reported more improvements in their communication and workflow. 

As future work, \klop{} is investigating the idea of extending the role of Slack as a central communication hub by adding more integrations with other tools, such as Confluence, and expanding the set of automation rules in Jira. Finally, they are considering also bringing to Berlin some of the changes investigated at the Italian site.

\bibliographystyle{ACM-Reference-Format}
\bibliography{biblio}
\end{document}